# Proximity effect and self-consistent field in a normal metal – superconductor structure

E.E. Zubov[1,2]

[1] G.V. Kurdyumov Institute for Metal Physics, NAS of Ukraine, 03680 Kyiv, Academician Vernadsky Boulevard, 36, Ukraine

[2] Vasyl' Stus Donetsk National University, 21021 Vinnytsia, str. 600-richchia, 21, Ukraine

E-mail: eezubov@ukr.net

The concept of a self-consistent field in the theory of superconductivity based on the diagram method of the time-dependent perturbation theory is presented. It is shown that the well-known BCS equation for the order parameter of superconductivity is already realized in a zero approximation. The form of interaction Hamiltonian uniquely determines a chain of interconnected Green's functions which are easily calculated in this approximation. On the basis of the presented method a proximity effect in a normal metal-superconductor structure is studied. It was obtained the energy gap values induced in a normal metal. In contrast to the traditional McMillan and de Gennes theories with self-consistent Green's functions the self-consistency over the order parameter gives a significantly smaller gap value induced in a normal metal. The frequency dependence of the homogeneous spectral density is obtained which qualitatively agrees with experiment.

Corresponding author: eezubov@ukr.net

## 1. Introduction

The study of the induced superconductivity in a normal metal adjacent to a superconductor has attracted the attention of many researchers for a long time [1–4]. It should be noted that despite the rather tremendous list of theoretical publications on this topic in most cases the main statements of the de Gennes [5] and McMillan [6] theories are used. As a rule, in this case for coordinate dependence of the gap an integral relation is realized which is directly connected with the well-known Gor'kov's equations [7]. If we consider the self-consistent field it is very difficult to extract it within the framework of these theories. The reason for this is the fact that self-consistency is always carried out by Green's functions and not by the parameter of order. For example, McMillan [6] uses the matrix form of the Green function proposed by Eliashberg [8] in the framework of the Nambu formalism for superconductors with a strong coupling and time delay effects. Here, in the framework of the Dyson equation the self-consistency is carried out for Green's functions also. A such approach significantly complicates the solution of the equations obtained and also leads to the necessity of inclusion the additional unknown parameters. In this case the induced gap is self-consistently connected with the energy gap of the superconductor. It is not well correctly since the role of the superconducting order parameter consists only in inducing a gap in a normal metal. As a result, the concept of the critical temperature $T_C$ of the superconducting transition arises for the whole hybrid structure. Although it is obvious that $T_C$ is determined solely by the value of the electron-phonon coupling parameter for superconductor. This leads to an overestimation of the induced energy gap value in the proximity effect. It should be noted that a such consideration leads to a need to take into account the spatial dependence of the energy gap function which satisfies the Eilenberger differential equation [9] with corresponding boundary conditions. The problem of the proximity effect in the case of a homogeneous gap that is quite natural for the ballistic limit remains open.

This paper proposes a new approach to solving the problem of superconductivity in both homogeneous and inhomogeneous structures. It is based on the use of the diagram method of time-dependent perturbation theory with selected order parameters according to which subsequent rigorous self-consistency is carried out. The advantages of a such theory are the presence of the minimum number of parameters used, a simpler form of the obtained equations for order parameters, as well as a clear knowledge of the realization one or another state of electron ensemble in a metal. Also, starting from the indicated approximation a subsequent account for the influence of fluctuations using the loop diagrams is possible.

The structure of the work is as follows: the second section presents the basic tenets of the self-consistent theory for normal metal; in the third section the BCS equation for the gap was first obtained by presented diagrammatic method. The derivation of this equation is based only on the graphical representation of the effective self-consistent field. In the fourth section the proximity effect in the normal metal-superconductor hybrid structure within the framework of a tunnel Hamiltonian is considered. The general expression for parameter of order is obtained and the temperature dependences of the induced superconducting energy gap in a normal metal are calculated. In the fifth section we study the spectral density of states of a normal metal with an induced energy gap function. In sixth section the main conclusions of the presented theory are given.

## 2. Self-consistent field in a normal metal. Diagrammatic method.

The Hamiltonian of a normal metal in a site representation can be written as

$$\hat{H} = \sum_{i,j,\sigma} t_{ij} c^+_{\sigma i} c_{\sigma j} - \mu \sum_{i\sigma} c^+_{\sigma i} c_{\sigma i} \quad , \tag{1}$$

where the hopping integral $t_{ij}$ determines the band energy $\varepsilon_{\mathbf{q}} = \sum_{ij} t_{ij} e^{-i\mathbf{q}(\mathbf{r}_i - \mathbf{r}_j)}$, $\mu$ and $\sigma$ are the chemical potential and electron spin, respectively. The creation $c^+_{i\sigma} = \frac{1}{\sqrt{N}} \sum_{k} e^{-ikR_i} c^+_{k\sigma}$ and annihilation $c_{i\sigma} = \frac{1}{\sqrt{N}} \sum_{k} e^{ikR_i} c_{k\sigma}$ operators are presented as the Fourier transforms in a wave space. It is convenient to take the part of Hamiltonian (1) associated with the chemical potential energy as an unperturbed Hamiltonian. The question immediately arises about the validity of a such choice. Strictly speaking the remaining part in (1) is not a perturbation for the metal. However, this choice is fully justified since the chemical potential in the ground state always limits the electron energy. Therefore, an infinite series of perturbation theory with

$$\hat{H}_0 = -\mu \sum_{i\sigma} c^+_{\sigma i} c_{\sigma i} \tag{2}$$

and perturbation Hamiltonian

$$V = \sum_{i,j,\sigma} t_{ij} c^+_{\sigma i} c_{\sigma j} \tag{3}$$

will be converging. Indeed, we associate the bold line with the total causal Green's function $Y^{-+}_{k\sigma}(\tau) = - <T c_{k\sigma}(\tau) c^+_{k\sigma}(0)>$ which is presented in fig.1. Here, the symbol $<...>$ denotes the statistical averaging over the full Hamiltonian (1), and $T_\tau$ is the time operator of chronological ordering, $\tau$ is the imaginary time in the Matsubara formalism.

A thin line corresponds to the unperturbed Green's function

$$G_{p\sigma}^{-+}(i\omega_n) = - <T_\tau c_{p\sigma}(\tau)c_{p\sigma}^+(0)>_{0,i\omega_n} = \frac{1}{\beta(i\omega_n + \mu)}, \qquad (4)$$

where a symbol $<...>_{0,i\omega_n}$ means the averaging with Hamiltonian $\hat{H}_0$. An imaginary index $i\omega_n = i\pi(2n+1)/\beta$ is the standard notation for the Fourier transform of Matsubara Green's function and $1/\beta = T$ is the temperature. The diagram in fig. 1 represents an infinite series of convergent geometric series with denominator $q = |\beta\varepsilon_p G_{p\sigma}^{-+}(i\omega_n)| < 1$. In the Fourier space

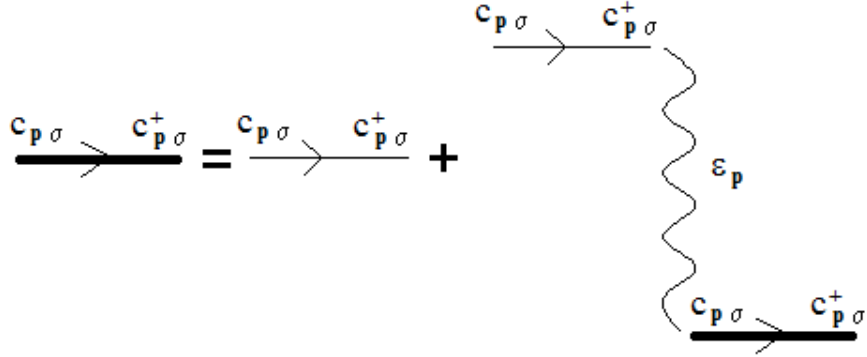

Fig.1. The complete diagram series for the Fourier transform of the causal Green's function $Y_{k\sigma}^{-+}(i\omega_n)$ in the zero approximation of the self-consistent field.

we have the following algebraic equation for unknown $Y_{k\sigma}^{-+}(i\omega_n)$:

$$Y_{k\sigma}^{-+}(i\omega_n) = G_{k\sigma}^{-+}(i\omega_n) + \beta\varepsilon_k Y_{k\sigma}^{-+}(i\omega_n), \qquad (5)$$

whose solution is trivial. In view of Eq.(4) it gives the traditional causal Green's function of the electron gas in a metal [10]:

$$Y_{k\sigma}^{-+}(i\omega_n) = \frac{1}{\beta(i\omega_n - \varepsilon_k + \mu)} \qquad (6)$$

Despite the triviality of deriving Eq.(6) for a zero-approximation of the self-consistent field this consideration has a deep physical meaning since it allows us to consider successively the contributions of correlation corrections related to the interaction of electrons and ions. As an example, in the next section we will consider the phenomenon of electon Cooper pairing in a system with electron-phonon coupling.

3. Self-consistent field in a superconducting metal

As is known, the basis for describing the phenomenon of superconductivity in a metal is the Bardeen-Cooper-Schrieffer Hamiltonian [11]:

$$V = \sum_{i,j,\sigma} t_{ij} a^+_{\sigma i} a_{\sigma j} - \sum_{i,j,\sigma} \left\{ \Delta_{ij\sigma} a^+_{\sigma i} a^+_{\sigma j} + \Delta^*_{ij\sigma} a_{-\sigma i} a_{\sigma j} \right\}, \qquad (7)$$

where for convenience the symbols of the creation and annihilation operators in the superconductor are denoted by the letter $a^+$ and $a$, respectively. In this case in the framework of suggested diagram method the Hamiltonian (7) is a perturbation. The unperturbed part of a total Hamiltonian over which all averaging of the perturbation theory series is performed has the form

$$\hat{H}_0 = -\mu \sum_{i\sigma} a^+_{\sigma i} a_{\sigma i} \qquad (8)$$

Fourier components $\Delta_{q\sigma}$ of the order parameters $\Delta_{ij\sigma}$ satisfy the self-consistency equation

$$\Delta_{k\sigma} = \sum_q \frac{|M_{k-q}|^2}{\omega_{k-q}} <a_{-q-\sigma} a_{q\sigma}>, \qquad (9)$$

where $M_{k-q}$ and $\omega_{k-q}$ are a matrix element of the electron-phonon interaction [10] and a phonon frequency, respectively. Thus, to calculate the order parameter it is necessary to know the expression for correlator $<a_{-q-\sigma} a_{q\sigma}>$. The classical expression for $\Delta_{k\sigma}$ known as the BCS equation was obtained [11] using the Bogolyubov's unitary transformation in a wave space. An elegant derivation of the BCS equation based on the equations of motion for Green's function was presented by Gorkov in [7].

We will show that a zero approximation of the self-consistent field also eventually leads to the BCS equation for order parameter. In full analogy with previous section it is necessary to present the graphic images of interactions of the perturbation Hamiltonian (7). In a wave representation the types of interaction lines are shown in fig.2. It turns out to be that two total Green's functions $Z^{--}_{q\sigma}(i\omega_n) = -<Ta_{-q-\sigma}(\tau) a_{q\sigma}(0)>_{i\omega_n}$ and $Z^{+-}_{q\sigma}(i\omega_n) = -<Ta^+_{q\sigma}(\tau) a_{q\sigma}(0)>_{i\omega_n}$ are now connected and it is necessary to take into account two unperturbed Green functions:

$$G^{-+}_{q\sigma}(i\omega_n) \equiv G^{-+}_{-q-\sigma}(i\omega_n) = -<Ta_{q\sigma}(\tau) a^+_{q\sigma}(0)>_{0,i\omega_n} = G_1(i\omega_n) = \frac{1}{\beta(i\omega_n + \mu)}$$
$$G^{+-}_{q\sigma}(i\omega_n) \equiv G^{+-}_{-q-\sigma}(i\omega_n) = -<Ta^+_{q\sigma}(\tau) a_{q\sigma}(0)>_{0,i\omega_n} = G_2(i\omega_n) = -G_1(-i\omega_n)$$
$$(10)$$

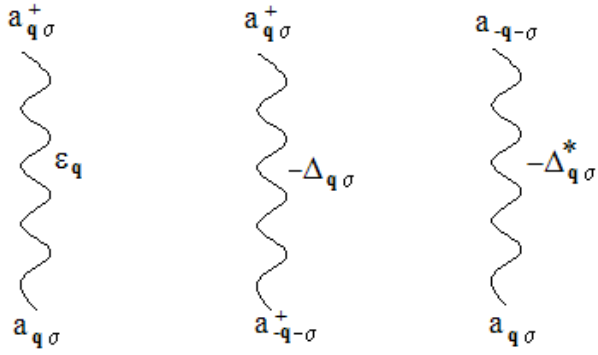

Fig.2. The all possible types of interaction lines for Hamiltonian (7).

It directly follows from the graphic equation in fig. 3 since a symbol of the end of Green's line must always coincide with a corresponding symbol at the origin of interaction line and vice versa (see Fig. 3). Also, the interaction line overturn changes a sign of the corresponding parameter to the opposite.

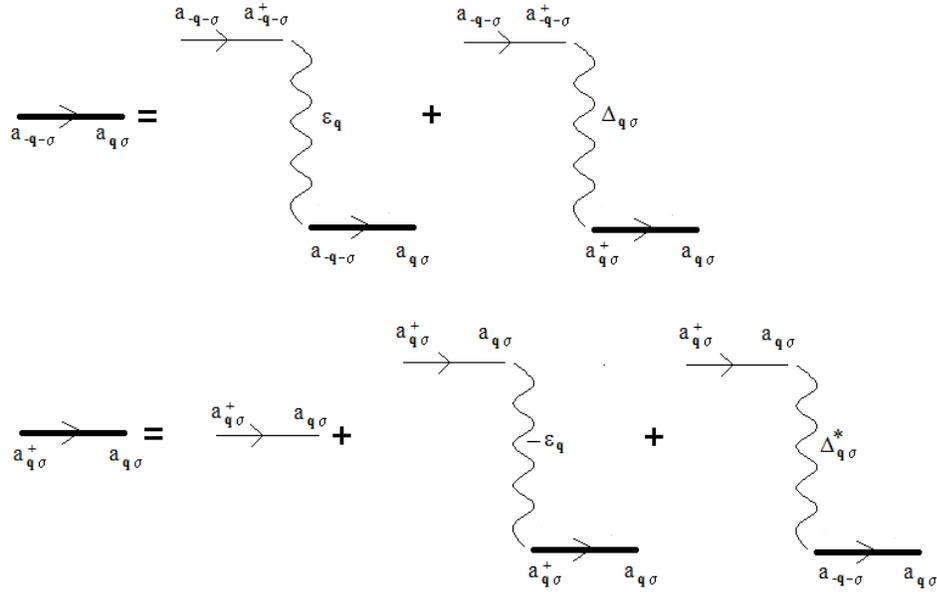

Fig. 3. The graphic system of equations for unknown total causal Green functions $Z^{--}_{q\sigma}(i\omega_n)$ and $Z^{+-}_{q\sigma}(i\omega_n)$.

The following system of algebraic equations corresponds to graphical system of equations in fig.3 for unknown terms:

$$Z^{--}_{-q-\sigma}(i\omega_n) = G_1(i\omega_n)\beta\Delta_{q\sigma}Z^{+-}_{q\sigma}(i\omega_n) + G_1(i\omega_n)\beta\varepsilon_{-q}Z^{--}_{-q-\sigma}(i\omega_n)$$
$$Z^{+-}_{q\sigma}(i\omega_n) = G_2(i\omega_n) - G_2(i\omega_n)\beta\varepsilon_q Z^{+-}_{q\sigma}(i\omega_n) + G_2(i\omega_n)\beta\Delta^*_{q\sigma}Z^{--}_{-q-\sigma}(i\omega_n)$$

(11)

It is not difficult to find a solution of this system with obvious condition $\varepsilon_q = \varepsilon_{-q}$. It can be written in the form:

$$Z_{q\sigma}^{+-}(i\omega_n) = \frac{G_2(i\omega_n)(1-\beta\varepsilon_q G_1(i\omega_n))((i\omega_n)^2 - \mu^2)}{(i\omega_n - E_{q\sigma})(i\omega_n + E_{q\sigma})}, \quad (12)$$

$$Z_{q\sigma}^{--}(i\omega_n) = \frac{\Delta_{q\sigma}}{2\beta E_{q\sigma}}\left\{\frac{1}{i\omega_n - E_{q\sigma}} - \frac{1}{i\omega_n - E_{q\sigma}}\right\}, \quad (13)$$

where $E_{q\sigma} = \sqrt{(\varepsilon_q - \mu)^2 + |\Delta_{q\sigma}|^2}$ is the electron excitation energy with Cooper pairing. The solution (12) determines a spectral density of excitations and infinitesimal corrections to the chemical potential. Therefore, below we will consider the solution (13) that gives the expression for correlator $<a_{-q-\sigma}a_{q\sigma}>$. It determines the self-consistent equation for the unknown gap function $\Delta_{q\sigma}$ according to Eq.(9). Indeed, for the times $\tau \to +0$ we have

$$<a_{-q-\sigma}a_{q\sigma}> = -Z_{-q-\sigma}^{--}(\tau \to +0) = <Ta_{-q-\sigma}(\tau \to +0)a_{q\sigma}(0)> = -\lim_{\tau \to +0}\sum_{n=-\infty}^{\infty}e^{-i\omega_n\tau}Z_{-q-\sigma}^{--}(i\omega_n) \quad (14)$$

Summation over an infinite number of frequencies is carried out on the basis of the method proposed by Luttinger and Ward in [12] for calculating the residues of the Green's function $Z_{-q-\sigma}^{--}(\omega)$ with a factor $\beta(f(\omega)-1)$, where $f(\omega) = \frac{1}{e^{\beta\omega}+1}$ is the Fermi distribution function. Thus, one can write

$$<a_{-q-\sigma}a_{q\sigma}> = -\beta\sum_i \text{Res}\left[Z_{-q-\sigma}^{--}(\omega)(f(\omega)-1)\right]\bigg|_i \quad (15)$$

From Eq.(13) it follows that the poles of $Z_{-q-\sigma}^{--}(\omega)$ are roots $\omega_i = \pm E_{q\sigma}$ that allows to immediately write for the correlator (15)

$$<a_{-q-\sigma}a_{q\sigma}> = -\frac{\Delta_{q\sigma}}{2E_{q\sigma}}\{f(E_{q\sigma}) - f(-E_{q\sigma})\} \quad (16)$$

Substituting Eq.(16) into Eq.(9), we obtain the well-known BCS equation for superconducting energy gap:

$$\Delta_{k\sigma} = \sum_q \frac{|M_{k-q}|^2}{\omega_{k-q}}\frac{\Delta_{q\sigma}}{2E_{q\sigma}}\tanh\left(\frac{E_{q\sigma}}{2T}\right), \quad (17)$$

where $\frac{|M_{k-q}|^2}{\omega_{k-q}}$ is a matrix element of the electron-electron attraction.

In this section it is obtained that a zero approximation of the self-consistent field is in fact an approximation of the molecular field in the theory of superconductivity. An influence of fluctuations can be taken into account by considering the contributions of loop diagrams in expressions for Green's functions. The presented theory allows us to consider the hybrid structure with normal metal and superconductor in which the induced superconductivity is realized as a phenomenon of proximity effect.

**4. Induced superconductivity in a normal metal**

Let us consider the hybrid structure with a normal metal and superconductor on the two sides of the junction. We assume that the connection between the metal and superconductor is carried out through a tunnel junction. Without applied voltage the two subsystems are in equilibrium with chemical potential $\mu$. It turns out to be the simplest approximation of the self-consistent field allows us to describe a proximity effect in this hybrid structure. It is supposed that the electronic subsystem of a normal metal can be in an ordered magnetic state. Then the unperturbed Hamiltonians of a normal "left" and superconducting "right" metals in the site representation have the following form, respectively:

$$\hat{H}_{0L} = -\sum_{i\sigma} \mu_\sigma c^+_{\sigma i} c_{\sigma i} \tag{18}$$

$$\hat{H}_{0R} = -\mu \sum_{i\sigma} a^+_{\sigma i} a_{\sigma i} , \tag{19}$$

where $\mu_\sigma = \mu + \sigma J_0$, $J_0$ is the exchange parameter and $J_0 > 0$ for a ferromagnetic metal. Here, $\sigma = \pm 1$ in the case of a saturated state and $\sigma = \pm 2 <\sigma_z>$ for a magnet with a mean spin $<\sigma_z>$.

The total perturbation Hamiltonian for a whole system is

$$V = \sum_{i,j,\sigma} t_{1ij} c^+_{\sigma i} c_{\sigma j} + \sum_{i,j,\sigma} t_{2ij} a^+_{\sigma i} a_{\sigma j} - \sum_{i,j,\sigma} \{\Delta_{ij\sigma} a^+_{\sigma i} a^+_{\sigma j} + \Delta^*_{ij\sigma} a_{-\sigma i} a_{\sigma j}\} + \hat{H}_T , \tag{20}$$

where the tunnel Hamiltonian is presented by the next manner

$$\hat{H}_T = \sum_{il\sigma} \{T_{il} c^+_{i\sigma} a_{l\sigma} + T^*_{il} a^+_{l\sigma} c_{i\sigma}\} \tag{21}$$

In full analogy with a diagram technique described in the previous sections let us draw graphically the possible types of interaction lines in a wave space in Fig.4. In this case one can assume that the wave vectors *p* and *q* with energies $\varepsilon_p$ and $\varepsilon_q$ belong to the "left" normal and "right" superconducting metals, respectively.

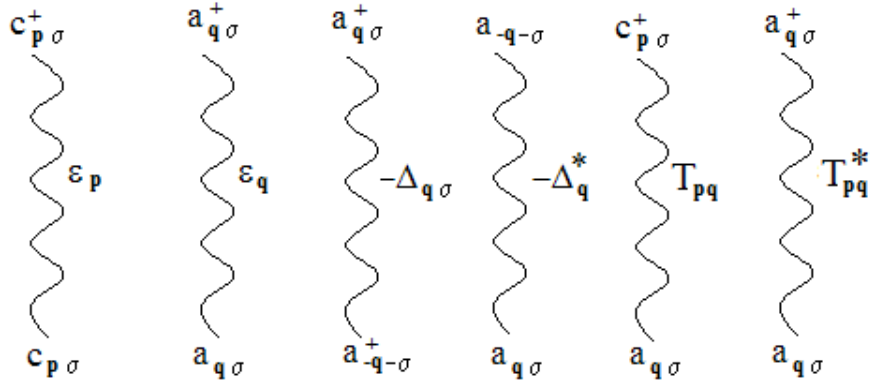

Fig.4. All possible types of interaction lines for Hamiltonian (20).

Here $T_{pq}$ and $T^*_{pq}$ are the matrix element and its complex conjugate value in a wave space, respectively. It is also necessary to consider that they depend on two wave vectors. Therefore, in the equations for Green's functions which depend only on a single wave vector the remaining one "internal" must be summed. Then one can write the graphical equations which determine the number of mutual connected Green's functions. In the presence of previously defined two unperturbed Green's functions $G_1(i\omega_n)$ and $G_2(i\omega_n)$ for superconducting "right" metal there are new functions

$$\tilde{G}^{-+}_{p\sigma}(i\omega_n) \equiv \tilde{G}^{-+}_{-p\sigma}(i\omega_n) = -<Tc_{p\sigma}(\tau)c^+_{p\sigma}(0)>_{0,i\omega_n} = \tilde{G}_{1\sigma}(i\omega_n) = \frac{1}{\beta(i\omega_n + \mu_\sigma)} \qquad (22)$$

$$\tilde{G}^{+-}_{p\sigma}(i\omega_n) \equiv \tilde{G}^{+-}_{-p\sigma}(i\omega_n) = -<Tc^+_{p\sigma}(\tau)c_{p\sigma}(0)>_{0,i\omega_n} = \tilde{G}_{2\sigma}(i\omega_n) = -G_{1\sigma}(-i\omega_n)$$

for part of the hybrid structure with a normal metal. Then the correlator $<c_{-p-\sigma}c_{p\sigma}>$ for "left" normal metal is connected with four complete Green's functions, i.e.

$$Y^{--}_{p\sigma}(\tau) = -<Tc_{-p-\sigma}(\tau)c_{p\sigma}(0)>, \quad Z^{--}_{qp\sigma}(\tau) = -<Ta_{-q-\sigma}(\tau)c_{p\sigma}(0)>, \quad Z^{+-}_{qp\sigma}(\tau) = -<Ta^+_{q\sigma}(\tau)c_{p\sigma}(0)>$$

and $\quad Y^{+-}_{p\sigma}(\tau) = -<Tc^+_{p\sigma}(\tau)c_{p\sigma}(0)>$.

Fig. 5 presents the graphical equations which are determined by Hamiltonian (20) with all possible lines of interaction in fig. 4. This equations connect the Fourier components of the indicated total Green's functions. Bold points in the diagrams denote a summation over the inner wave vector.

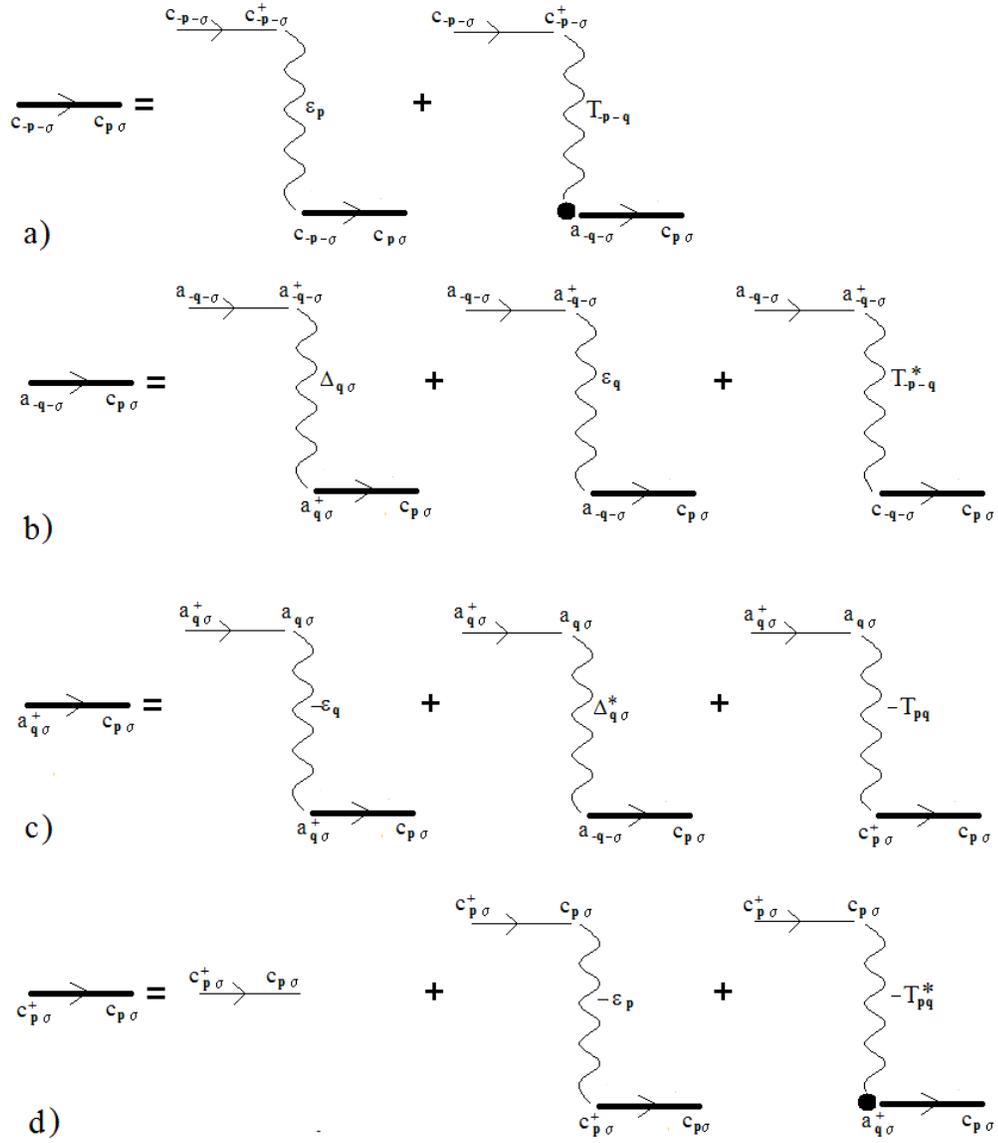

Fig. 5. The graphic system of equations for unknown total causal Green's functions $Y^{--}_{p\sigma}(i\omega_n)$, $Z^{--}_{qp\sigma}(i\omega_n)$, $Z^{+-}_{qp\sigma}(i\omega_n)$ и $Y^{+-}_{p\sigma}(i\omega_n)$.

One can write these graphical system of equations in algebraic form:

$$Y^{--}_{p\sigma}(i\omega_n) = \tilde{G}_{1-\sigma}\beta\varepsilon_p(i\omega_n)Y^{--}_{p\sigma}(i\omega_n) + \tilde{G}_{1-\sigma}\sum_q T_{-p-q}Z^{--}_{qp\sigma}(i\omega_n)$$
$$Z^{--}_{qp\sigma}(i\omega_n) = G_1(i\omega_n)\beta\Delta_{q\sigma}Z^{+-}_{qp\sigma}(i\omega_n) + G_1(i\omega_n)\beta\varepsilon_q Z^{--}_{qp\sigma}(i\omega_n) + G_1(i\omega_n)\beta T^*_{-p-q}Y^{--}_{p\sigma}(i\omega_n)$$
$$Z^{+-}_{qp\sigma}(i\omega_n) = -G_2(i\omega_n)\beta\varepsilon_q Z^{+-}_{qp\sigma}(i\omega_n) + G_2(i\omega_n)\beta\Delta^*_{q\sigma}Z^{--}_{qp\sigma}(i\omega_n) - G_2(i\omega_n)\beta T_{pq}Y^{+-}_{p\sigma}(i\omega_n)$$
$$Y^{+-}_{p\sigma}(i\omega_n) = \tilde{G}_{2\sigma}(i\omega_n) - \tilde{G}_{2\sigma}(i\omega_n)Y^{+-}_{p\sigma}(i\omega_n) - \tilde{G}_{2\sigma}(i\omega_n)\sum_q \beta T^*_{pq}Z^{+-}_{qp\sigma}(i\omega_n)$$

(23)

In spite of the fact that this system of equations is integral it is easy to find its solution. Indeed, the 2-nd and 3-rd equations are linear with respect to the Green's functions depending simultaneously on the *p* and *q* wave vectors. Thus, expressing the off-diagonal $Z^{--}_{qp\sigma}(i\omega_n)$ and $Z^{+-}_{qp\sigma}(i\omega_n)$ in terms of diagonal we have

$$Z^{--}_{qp\sigma}(i\omega_n) = \frac{\beta G_1(i\omega_n)}{d_{q\sigma}(i\omega_n)} \left\{ T^*_{-p-q}(1+\beta\varepsilon_q G_2(i\omega_n))Y^{--}_{p\sigma}(i\omega_n) - T_{pq}\beta\Delta_{q\sigma}G_2(i\omega_n)Y^{+-}_{p\sigma}(i\omega_n) \right\}$$
$$Z^{+-}_{qp\sigma}(i\omega_n) = \frac{\beta G_2(i\omega_n)}{d_{q\sigma}(i\omega_n)} \left\{ -T_{pq}(1-\beta\varepsilon_q G_1(i\omega_n))Y^{+-}_{p\sigma}(i\omega_n) + T^*_{-p-q}\beta\Delta^*_{q\sigma}G_1(i\omega_n)Y^{--}_{p\sigma}(i\omega_n) \right\}$$
(24)

where

$$d_{q\sigma}(i\omega_n) = (1-\beta\varepsilon_q G_1(i\omega_n))(1+\beta\varepsilon_q G_2(i\omega_n)) - \beta^2 |\Delta_{q\sigma}|^2 G_1(i\omega_n)G_2(i\omega_n) \quad (25)$$

Substituting the solutions (24) into the 1-st and 4-th equations of system (23) we obtain a system of two linear equations for the unknowns $Y^{--}_{p\sigma}(i\omega_n)$ and $Y^{+-}_{p\sigma}(i\omega_n)$ which do not depend on the summation over the wave vector *q*. For convenience we put the following notation

$$\alpha_{p\sigma}(i\omega_n) = \beta^2 \tilde{G}_{1-\sigma}(i\omega_n)G_1(i\omega_n)\sum_q \frac{1}{d_{q\sigma}(i\omega_n)} |T_{pq}|^2 (1+\beta\varepsilon_q G_2(i\omega_n))$$

$$\beta_{p\sigma}(i\omega_n) = \beta^3 \tilde{G}_{1-\sigma}(i\omega_n)G_1(i\omega_n)G_2(i\omega_n)\sum_q \frac{1}{d_{q\sigma}(i\omega_n)} |T_{pq}|^2 \Delta_{q\sigma}$$

$$\gamma_{p\sigma}(i\omega_n) = \beta^2 \tilde{G}_{2\sigma}(i\omega_n)G_2(i\omega_n)\sum_q \frac{1}{d_{q\sigma}(i\omega_n)} |T_{pq}|^2 (1-\beta\varepsilon_q G_1(i\omega_n))$$

$$\theta_{p\sigma}(i\omega_n) = \beta^3 \tilde{G}_{2\sigma}(i\omega_n)G_1(i\omega_n)G_2(i\omega_n)\sum_q \frac{1}{d_{q\sigma}(i\omega_n)} |T_{pq}|^2 \Delta^*_{q\sigma}$$
(26)

Here, the identity $T^*_{pq} = T_{-p-q}$ for the tunnel matrix element is used. Then the above system of linear equations takes the form:

$$Y^{--}_{p\sigma}(i\omega_n) = \left(\tilde{G}_{1-\sigma}(i\omega_n)\beta\varepsilon_p + \alpha_{p\sigma}(i\omega_n)\right)Y^{--}_{p\sigma}(i\omega_n) - \beta_{p\sigma}(i\omega_n)Y^{+-}_{p\sigma}(i\omega_n)$$
$$Y^{+-}_{p\sigma}(i\omega_n) = \tilde{G}_{2\sigma}(i\omega_n) - \theta_{p\sigma}(i\omega_n)Y^{--}_{p\sigma}(i\omega_n) + (\gamma_{p\sigma}(i\omega_n) - \beta\varepsilon_p\tilde{G}_{2\sigma}(i\omega_n))Y^{+-}_{p\sigma}(i\omega_n)$$
(27)

The solution of this system of equations is written as follows:

$$Y^{+-}_{p\sigma}(i\omega_n) = -\frac{1 - \beta\varepsilon_p\tilde{G}_{1-\sigma}(i\omega_n) - \alpha_{p\sigma}(i\omega_n)}{\beta_{p\sigma}(i\omega_n)} Y^{--}_{p\sigma}(i\omega_n)$$

$$Y^{--}_{p\sigma}(i\omega_n) = \frac{\tilde{G}_{2\sigma}(i\omega_n)\beta_{p\sigma}(i\omega_n)}{\theta_{p\sigma}(i\omega_n)\beta_{p\sigma}(i\omega_n) - \left[1 + \beta\varepsilon_p\tilde{G}_{2\sigma}(i\omega_n) - \gamma_{p\sigma}(i\omega_n)\right]\left[1 - \beta\varepsilon_p\tilde{G}_{1-\sigma}(i\omega_n) - \alpha_{p\sigma}(i\omega_n)\right]}$$
(28)

We are interested in the Green's function $Y^{--}_{p\sigma}(i\omega_n)$ since it defines the energy gap in a normal metal. In accordance with Eq. (15) the real frequencies are used to calculate the correlators. Then in all functions we make a replacement $i\omega_n \to \omega$. Substituting the unperturbed Green's functions $G_1(\omega)$ and $G_2(\omega)$ in Eq. (25) it is easy to find

$$d_{q\sigma}(\omega) = \frac{\omega^2 - \mu^2}{(\omega - E_{q\sigma})(\omega + E_{q\sigma})}, \tag{29}$$

where $E_{q\sigma}$ is the electron excitation energy from Eq.(13) for a superconductor. Similarly, for real frequencies we have

$$\alpha_{p\sigma}(\omega) = \frac{\tilde{\alpha}_{p\sigma}}{\omega + \mu_{-\sigma}}, \quad \beta_{p\sigma}(\omega) = \frac{\tilde{\beta}_{p\sigma}}{\omega + \mu_{-\sigma}}, \quad \gamma_{p\sigma}(\omega) = \frac{\tilde{\gamma}_{p\sigma}}{\omega - \mu_{\sigma}}, \quad \theta_{p\sigma}(\omega) = \frac{\tilde{\theta}_{p\sigma}}{\omega - \mu_{\sigma}}, \tag{30}$$

where $\xi_q = \varepsilon_q - \mu$ and

$$\begin{aligned}
\tilde{\alpha}_{p\sigma}(\omega) &= \sum_q |T_{pq}|^2 \frac{\omega + \xi_q}{(\omega - E_{q\sigma})(\omega + E_{q\sigma})} \\
\tilde{\beta}_{p\sigma}(\omega) &= \sum_q |T_{pq}|^2 \Delta_{q\sigma} \frac{1}{(\omega - E_{q\sigma})(\omega + E_{q\sigma})} \\
\tilde{\gamma}_{p\sigma}(\omega) &= \sum_q |T_{pq}|^2 \frac{\omega - \xi_q}{(\omega - E_{q\sigma})(\omega + E_{q\sigma})} \\
\tilde{\theta}_{p\sigma}(\omega) &= \tilde{\beta}^*_{p\sigma}(\omega)
\end{aligned} \tag{31}$$

The Green's function $Y^{--}_{p\sigma}(i\omega_n)$ is determined as

$$Y^{--}_{p\sigma}(\omega) = -\frac{1}{\beta} \frac{\tilde{\beta}_{p\sigma}(\omega)}{(\omega + \xi_p - J_0\sigma - \tilde{\gamma}_{p\sigma}(\omega))(\omega - \xi_p - J_0\sigma - \tilde{\alpha}_{p\sigma}(\omega)) - |\tilde{\beta}_{p\sigma}(\omega)|^2} \tag{32}$$

One can see from Eq.(32) to find the $Y^{--}_{p\sigma}(\omega)$ poles it is necessary to solve the complicated integral equations with a frequency $\omega$. The problem is significantly simplified if we assume that an absolute value of the tunnel matrix element square $|T_{pq}|^2$ is much less than the electron excitation energies $\omega_{01} = \xi_p + J_0\sigma$ and $\omega_{02} = -\xi_p + J_0\sigma$ in a normal metal without tunneling. Then in sums of Eqs. (31) the frequencies $\omega$ can be replaced by corresponding $\omega_{0i}$ and Green's function poles (32) are written in the form:

$$\begin{aligned}
\omega_1 &= \omega_{01} + \tilde{\alpha}_{p\sigma}(\omega_{01}) \\
\omega_2 &= \omega_{02} + \tilde{\gamma}_{p\sigma}(\omega_{02})
\end{aligned}, \tag{33}$$

where $|\tilde{\beta}_{p\sigma}(\omega)|^2 \Box |T_{p\sigma}|^4$ is taken into account. In Eq. (32) this contribution to $Y^{--}_{p\sigma}(\omega)$ poles is neglected. It should be noted that in another extreme case $|T_{pq}| >> \omega_{0i}$ with obvious condition $\mu >> |T_{pq}|$ we have $\tilde{\beta}_{p\sigma}(\omega) \sim |T_{p\sigma}|^2 \Delta \rho(\mu) / \mu^{3/2} \sim 0$ where $\rho(\mu)$ is the free electron density of states at the Fermi level (see Appendix, formula (A.1)). Thus, with a high barrier transparency the Cooper pairs of a superconductor are destroyed in a normal metal and the proximity effect is not observed.

Now one can write the final expression for $Y^{--}_{p\sigma}(\omega)$

$$Y^{--}_{p\sigma}(\omega) = \frac{1}{\beta}\left\{\frac{\tilde{\beta}_{p\sigma}(\omega_{02})}{(2\xi_p + \tilde{\alpha}_{p\sigma}(\omega_{02}) - \tilde{\gamma}_{p\sigma}(\omega_{02}))(\omega - \omega_2)} - \frac{\tilde{\beta}_{p\sigma}(\omega_{01})}{(2\xi_p + \tilde{\alpha}_{p\sigma}(\omega_{01}) - \tilde{\gamma}_{p\sigma}(\omega_{01}))(\omega - \omega_1)}\right\} \quad (34)$$

Further we consider the ballistic limit when $|T_{p\sigma}| = B$. Also, we put $\Delta_{q\sigma} = \Delta$, i.e. the superconducting energy gap is real and nondispersive. Then we have

$$\begin{aligned}\tilde{\alpha}_{p\sigma}(\omega_{0i}) - \tilde{\gamma}_{p\sigma}(\omega_{0i}) &= 2B^2\chi(\omega_{0i}) \\ \tilde{\beta}_{p\sigma}(\omega_{0i}) &= B^2\Delta\varphi(\omega_{0i})\end{aligned}, \quad (35)$$

Where

$$\varphi(\omega) = \sum_q \frac{1}{(\omega^2 - \xi_q^2 - \Delta^2)}$$

$$\chi(\omega) = \sum_q \frac{\xi_q}{(\omega^2 - \xi_q^2 - \Delta^2)} \quad (36)$$

It is obvious that correlator $<c_{-p-\sigma}c_{p\sigma}>$ of the electron singlet pairing in a normal metal is expressed in terms of functions $\varphi(\omega_{0i})$ and $\chi(\omega_{0i})$ using the formula

$$<c_{-p-\sigma}c_{p\sigma}> = -\beta\sum_i \mathrm{Res}\left[Y^{--}_{p\sigma}(\omega)(f(\omega)-1)\right]\bigg|_i, \quad (37)$$

as in the the previous section. The result is

$$<c_{-p-\sigma}c_{p\sigma}> = -\frac{1}{2}B^2\Delta\left\{\frac{\varphi(\omega_{02})[f(\omega_2)-1]}{\xi_p + B^2\chi(\omega_{02})} - \frac{\varphi(\omega_{01})[f(\omega_1)-1]}{\xi_p + B^2\chi(\omega_{01})}\right\} \quad (38)$$

This correlator produces an energy gap function of induced superconductivity which is determined by the matrix element $\tilde{V}_{kp}$ of electron-electron attraction in a normal metal by means of the standard equation

$$\tilde{\Delta}_{k\sigma} = \sum_p \tilde{V}_{kp} <c_{-p-\sigma}c_{p\sigma}> \quad (39)$$

Since the "left" metal is in the normal state the Eq.(39) is not self-consistent with respect to the order parameter $\tilde{\Delta}_{k\sigma}$. This parameter is induced exclusively by a gap function $\Delta$ of the "right" superconducting metal.

As can be seen from Eq.(38), for finding $<c_{-p-\sigma}c_{p\sigma}>$ it is necessary to calculate the integrals (36) that is not difficult. We will neglect the exchange interaction of electrons, i.e.

$J_0 = 0$. For a metal the obvious relation $\mu >> \Delta$ is satisfied. The results of calculating the integrals (36) are given in the Appendix. In particular, it is obtained that at $|\Delta| < |\xi_p|$ the function $\varphi(\pm \xi_p) = 0$, i.e. in a normal metal the electrons with an energy $\xi_p$ that exceeds the superconducting energy gap do not pair. In this case the proximity effect in a hybrid structure is not realized. At $|\Delta| > |\xi_p|$ the expressions for $\varphi(\xi_p)$ and $\chi(\xi_p)$ take the form (see (A.7) and (A.8)):

$$\varphi(\xi_p) = -\frac{\pi \rho(\mu)}{2\sqrt{\Delta^2 - \xi_p^2}}$$

$$\chi(\xi_p) = -\rho(\mu)\left\{2 + \ln \frac{\sqrt{\Delta^2 - \xi_p^2}}{4\mu}\right\} \quad (40)$$

Substituting functions (40) into Eq.(38) and using Eq.(39) we obtain the equation for induced gap in a normal metal:

$$\tilde{\Delta} = \frac{1}{8}\pi \tilde{\lambda} r \Delta \int_{-\Delta}^{\Delta} d\xi \frac{\tanh\left(\frac{\xi}{2T}\right)}{\sqrt{\Delta^2 - \xi^2}\{\xi + c\}} \quad (41)$$

where $c = -r\left[2 + \ln \frac{\sqrt{\Delta^2 - \xi^2}}{4\mu}\right]$, $r = 2B^2 \rho(\mu)$ and $\tilde{\lambda} = \tilde{V}_{kp}\tilde{\rho}(\mu)$ is the constant of effective electron attraction in a normal metal with density of state $\tilde{\rho}(\mu)$ on the Fermi level. In the McMillan notation [6] for a superconductor unit volume the quantity $r = 2\Gamma_N$ is determined by the relaxation time in a normal state. We suppose that the homogeneity condition $\tilde{\Delta}_{k\sigma} = \tilde{\Delta}$ is satisfied for induced gap in the ballistic limit.

The integral in Eq.(41) contains a pole singularity at $\xi = -c < 0$. It indicates the need to take into account for adiabatic inclusion of the interaction associated with the tunnel Hamiltonian. Therefore, the integrand in Eq.(41) is analytically continued to the upper complex plain that makes it possible to write integral (41) in the form

$$\int_{-\Delta}^{\Delta} \ldots d\xi = V.p. \int_{-\Delta}^{\Delta} \ldots d\xi - i\pi \int_{-\Delta}^{\Delta} \ldots \delta(\xi + c)d\xi \quad (42)$$

in accordance with the Landau bypass rule [1]. Here, $V.p.$ denotes a principal real value of the integral and $\delta(x)$ is the delta Dirac function which determines its imaginary part. It is also necessary to take into account that at $T = 0$ we have $\tanh\left(\dfrac{\xi}{2T}\right) = sign(\xi)$. It is supposed that $B \ll \Delta$ and then

$$c \approx \tilde{c}(r,\Delta) = -r\left[2 + \ln\dfrac{\Delta}{4\mu}\right], \qquad (43)$$

since the corrections will be of a higher order of smallness over the parameter $B$. Also, $c > 0$ due to the ratio $4\mu \gg \Delta$. It should be noted that in the most cases $\ln\dfrac{\Delta}{4\mu} \sim -10$ that gives for $c \sim 8r$. Therefore, at least the inequality $\Delta > 8r$ is satisfied. Numerically integrating (41) as the main value we obtain a dependence of the real part of the induced gap $\tilde{\Delta}$ on both temperature $T$ and $\Delta$ for different values of parameters. The imaginary part of this integral determines the phase $\psi$ of the induced order parameter. At temperature $T = 0$ it is easy to find the analytical expressions for both $\text{Re}(\tilde{\Delta})$ and $\psi$ which can be written as

$$\text{Re}(\tilde{\Delta}) = \dfrac{1}{8}\pi\tilde{\lambda}r\dfrac{\Delta}{\sqrt{\Delta^2 - \tilde{c}(r,\Delta)^2}}P(\Delta,r)$$

$$\psi = \arctan\dfrac{\pi}{P(\Delta,r)} \qquad (44)$$

where $P(\Delta,r) = \ln\dfrac{\Delta + \sqrt{\Delta^2 - \tilde{c}(r,\Delta)^2}}{\Delta - \sqrt{\Delta^2 - \tilde{c}(r,\Delta)^2}}$. The absolute value of the induced gap has the form

$$\left|\tilde{\Delta}\right| = \dfrac{1}{8}\pi\tilde{\lambda}r\dfrac{\Delta}{\sqrt{\Delta^2 - \tilde{c}(r,\Delta)^2}}\sqrt{P^2(\Delta,r) + \pi^2} \qquad (45)$$

It can be seen from Eqs.(44) that with $r \to 0$ the both gap $\tilde{\Delta}$ and phase $\psi$ also tend to zero despite the divergence of the natural logarithm. In fig. 6 *a* and *b* the dependences of the relative gap $\left|\tilde{\Delta}/\Delta\right|$ and induced $\tilde{\Delta}$ on the value *r* and superconducting gap $\Delta$ at zero temperature for $\mu = 5$ eV and $\tilde{\lambda} = 0.14$ are presented, respectively. The figure shows that the induced gap value can be amount to 5-6% percent of the original although in the area $\Delta \sim 6r$ the observed strong growth is associated with an error of the chosen approximation when $\Delta \sim \tilde{c}(r,\Delta)$. With an increase the probability of tunneling the proximity effect is strengthen as well as with an increase the superconducting gap value a relative value of the induced $\tilde{\Delta}$ decreases.

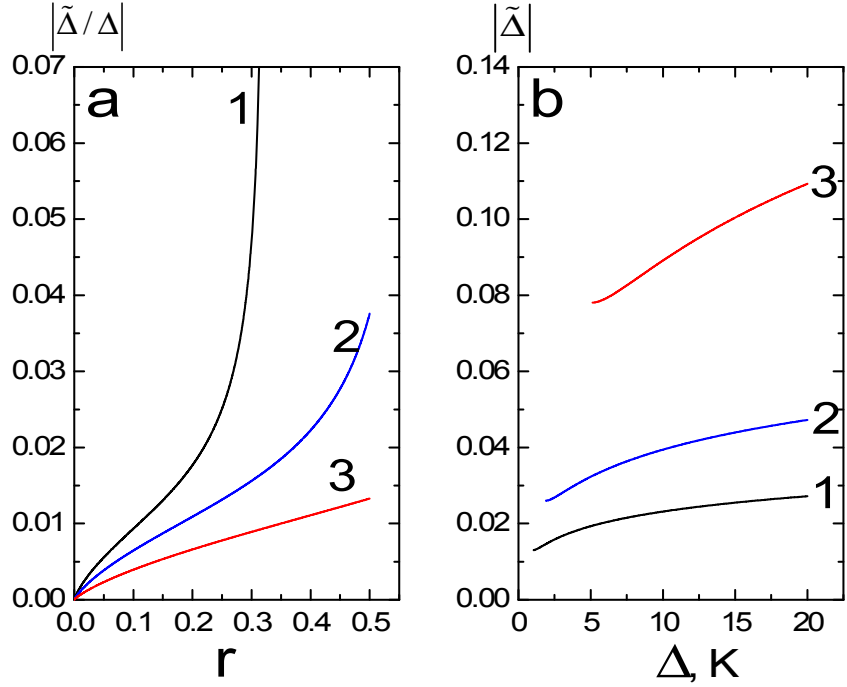

Fig. 6. The dependencies at temperature $T = 0$ a) the absolute value of relative gap $\left|\tilde{\Delta}/\Delta\right|$ on the doubled barrier transparency $r = 2\Gamma_N$ at $\Delta = 3$, 5 and 10 K and b) the induced gap $\tilde{\Delta}$ on the superconducting gap $\Delta$ at $r = 0.05$, 0.1 and 0.3 K (curves 1-3, respectively). The values $\mu = 5$ eV and $\tilde{\lambda} = 0.14$ were used.

The phase of induced order parameter depend on both $r$ and $\Delta$. Fig. 7 presents the dependences of induced order parameter phase on the doubled transparency $r$ (*a*) and the superconducting energy gap function $\Delta$ (*b*). From the figure it can be seen that for fixed $\Delta$ the phase of $\tilde{\Delta}$ increases from zero to 90 ° with increase a barrier transparency. For a given barrier transparency an increase in $\Delta$ leads to a decrease in the phase which for a large $\Delta$ slowly approaches to zero in accordance with the limit $\Delta \to \infty$ in Eq.(44) for $\psi$.

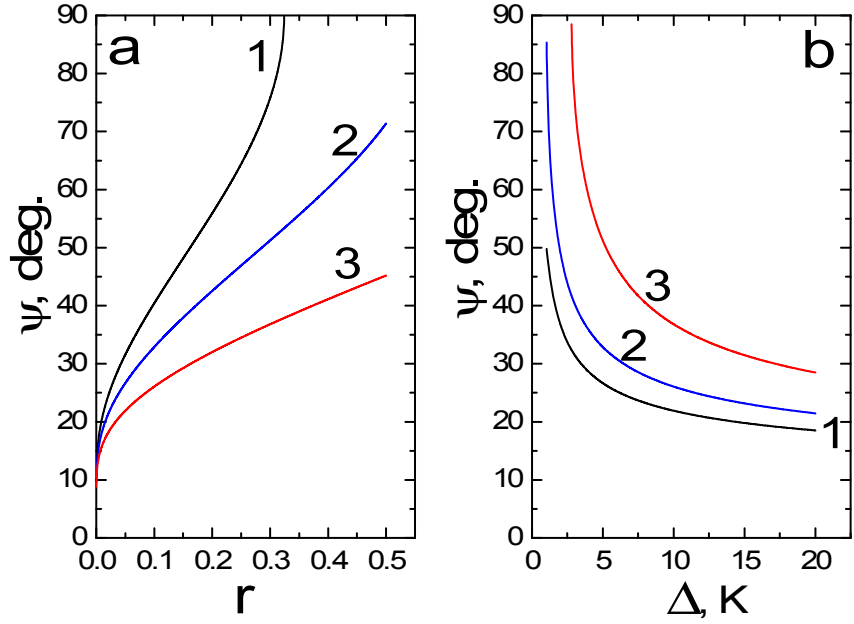

Fig. 7. The dependences of order parameter phase $\psi$ at temperature $T = 0$ a) on the doubled barrier transparency $r = 2\Gamma_N$ at $\Delta = 3$, 5 and 10 K and b) on the superconducting energy gap function $\Delta$ at $r = 0.05$, 0.1 and 0.3 K (curves 1- 3, respectively) with chemical potential $\mu = 5$ eV.

In conclusion we will consider the temperature dependences of absolute value of induced superconducting energy gap and its phase in the hybrid structure Ti-barrier-Sn. The critical temperatures of phase transitions in the superconducting state for Ti and Sn are 0.4 and 3.75 K, respectively. The constants of electron effective attraction in normal metals Ti and Sn are 0.141 and 0.245 [13], respectively. Fig. 8 presents the temperature dependence of energy gap function $\Delta$ for Sn which we will use to calculate $\tilde{\Delta}$ as a function of $T$ using Eq. (41). To calculate $\Delta(T)$ the relation $\omega_D(\lambda) = 0.5\Delta_0 \exp(1/\lambda)$ for the Debye frequency is taken into account, where $\Delta_0 = 6.6$ K is the superconducting energy gap for Sn at $T=0$. In this case the root of equation $\xi=c$ is determined numerically but not by the approximate formula (43). Then from Eq.(41) it is easy to find

$$|\tilde{\Delta}| = \sqrt{\left[\operatorname{Re}(\tilde{\Delta})\right]^2 + \left[\operatorname{Im}(\tilde{\Delta})\right]^2} \ , \tag{46}$$

where $\text{Im}(\tilde{\Delta}) = \dfrac{\pi^2 \tilde{\lambda} r \Delta \tanh\left(\dfrac{c}{2T}\right)}{8\sqrt{\Delta^2 - c^2}}$. Obviously, the phase $\psi$ of the induced energy gap has the form

$$\psi = \arctan\left(\frac{\text{Im}(\tilde{\Delta})}{\text{Re}(\tilde{\Delta})}\right) \qquad (47)$$

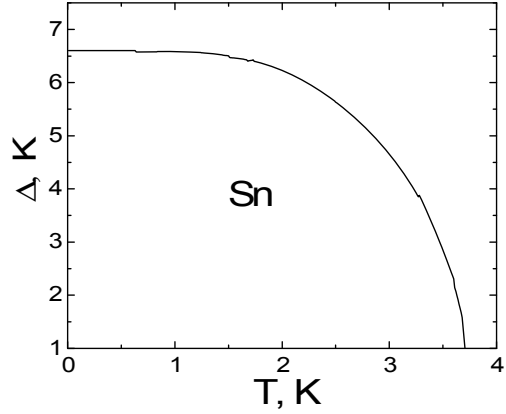

Fig.8. Temperature dependence of the superconducting energy gap $\Delta$ for Sn.

In fig. 9 the temperature dependences of the both $|\tilde{\Delta}|$ and $\psi$ for energy gap induced in Ti are presented. From fig.9 it folllows that for $r = 0.3$ K the maximum proximity effect is only a few percent of the original gap. In the area of a $\Delta$ strong decay the calculation becomes difficult due to violation of the basic requirement $\Delta > 8r$. Note that with increasing the temperature the order parameter phase is non-monotonic and oscillates around its value at zero temperature.

Thus, in the framework of a weak barrier transparency approximation it was obtained that the proximity effect is only a few percent of the initial superconducting energy gap. In the case of strong barrier transparency the tunneling destroys the Cooper pairs in a normal metal that leads to disappearance of the proximity effect.

## 5. The spectral density of states in a normal metal – superconductor structure

It is interesting to find the spectral density of states which is determined experimentally by differential conductivity [10]. It is known [10] the Green's functions to be obtained in the previous sections (see Eqs.(12) and (28)) have imaginary parts determined by means of analytic continuation $\omega \to \omega + i\delta$. In this case the general property $G_{q\sigma}^{-+}(i\omega_n) = -G_{q\sigma}^{+-}(-i\omega_n)$ is used. Then for a superconductor one can write the spectral density

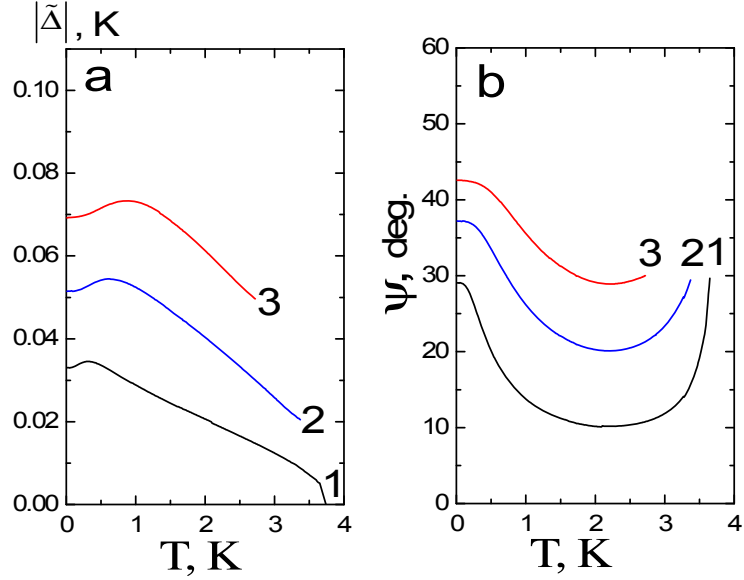

Fig. 9. The temperature dependences of a) absolute value of the induced superconducting energy gap and b) the phase $\psi$ of $\tilde{\Delta}$ in the hybrid structure Ti-barrier-Sn for double transparency values $r= 0.1, 0.2$ and $0.3$ K (curves 1-3, respectively). The chemical potential is assumed to be 5 eV.

$$A_{q\sigma}(\omega) = -2\beta \operatorname{Im} Z_{q\sigma}^{-+}(\omega + i\delta), \tag{48}$$

and for normal metal in the considered hybrid structure

$$A_{p\sigma}(\omega) = -2\beta \operatorname{Im} Y_{q\sigma}^{-+}(\omega + i\delta) \tag{49}$$

Further we are interested in homogeneous spectral densities $R_\sigma^S(\omega) = \sum_q A_{q\sigma}(\omega)$ and $R_\sigma^N(\omega) = \sum_p A_{p\sigma}(\omega)$ for both the superconductor and metal, respectively. From Eq.(12) at $\mu \gg \omega$ it is easy to find the traditional spectral density of superconductor at $\omega > \Delta$:

$$R_\sigma^S(\omega) = \frac{\pi \rho(\mu) \omega}{\sqrt{\omega^2 - \Delta^2}} \tag{50}$$

For a normal metal from Eq.(28) in the linear approximation with $\Gamma_N$ it follows that

$$Y_{p\sigma}^{-+}(\omega) = \frac{G_1(\omega)}{1 - \beta \varepsilon_p G_1(\omega) - \gamma_{p\sigma}(-\omega)}$$

Taking into account Eqs. (31) we have for retarded Green's function which determines the spectral density of states

$$Y_{p\sigma}^{-+}(\omega+i\delta) = \frac{1}{\beta} \frac{1}{\omega+i\delta - \xi_p - \tilde{\alpha}_{p\sigma}(\omega+i\delta)}, \tag{51}$$

where $\tilde{\alpha}_{p\sigma}(\omega) = B^2(\omega\varphi(\omega) + \chi(\omega))$ and functions $\varphi(\omega)$ and $\chi(\omega)$ are defined by Eqs.(36). From Eqs.(36) it follows that the analytic continuation of functions $\varphi(\omega)$ and $\chi(\omega)$ gives the imaginary parts only for $\omega > \Delta$ because of the delta functions presence. Thus, for excitation energies $\omega < \Delta$, i.e. in area of the proximity effect realization, the spectrum of electronic excitations is coherent. From (A.5) and (A.7) at $\sqrt{\Delta^2 - \omega^2} \ll \mu$ it follows that

$$\varphi(\omega) \approx -\frac{\rho(\mu)}{\sqrt{\Delta^2 - \omega^2}} \arctan\left(\frac{2\mu}{\sqrt{\Delta^2 - \omega^2}}\right)$$

$$\chi(\omega) \approx -\rho(\mu)\left\{2 + \ln\left(\frac{\sqrt{\Delta^2 - \omega^2}}{4\mu}\right)\right\} \tag{52}$$

Obviously, $\tilde{\alpha}_{p\sigma}(\omega)$ is a mass operator for Green's function $Y_{p\sigma}^{-+}(\omega)$ [10]. Therefore, one can write the spectral density of coherent excitations in a normal metal as

$$A_{p\sigma}(\omega) = 2\pi Z(\omega)\delta(\omega - \xi_p), \tag{53}$$

where the intensity of a quasiparticle peak is of the form

$$Z_{p\sigma}(\omega) = \frac{1}{\left|1 - \frac{\partial}{\partial \omega}\tilde{\alpha}_{p\sigma}(\omega)\right|_{\omega=\xi_p}} \tag{54}$$

and

$$\frac{\partial}{\partial \omega}\tilde{\alpha}_{p\sigma}(\omega) = B^2(\varphi(\omega) + \omega\varphi'(\omega) + \chi'(\omega)) \tag{55}$$

Using Eqs.(52) we calculate the derivatives $\varphi'(\omega)$ and $\chi'(\omega)$ and substitute in Eq.(55) and further in Eqs.(54) - (53). The sum of the functions $A_{p\sigma}(\omega)$ over vectors gives the final expression for the frequency dependence of the uniform spectral density

$$R_\sigma^N(\omega) = \frac{2\pi\rho(\mu)}{\left|1 - \Gamma_N\left\{\frac{\omega}{\Delta^2 - \omega^2} - \frac{\pi\Delta^2}{2\sqrt{(\Delta^2 - \omega^2)^3}}\right\}\right|} \tag{56}$$

When $\Gamma_N = 0$ we obtain the spectral density of normal metal. It can be seen from Eq.(56) that with decrease $\Gamma_N$ the spectral density increases approaching a constant value $2\pi\rho(\mu)$ for a normal metal. Thus, the proximity effect weakens. If $\Gamma_N$ increases the contribution of the unit in the denominator Eq.(56) decreases and $R_\sigma^N(\omega) \sim 1/\Gamma_N$, i.e. the spectral density decreases to zero. In this case the normal metal approaches the superconducting state. Note that in area $\omega<\Delta$ of coherent excitations the spectral density has a maximum the position of which is determined only by the superconducting energy gap $\Delta$. Indeed, the extremum point $\omega=\omega_{ext}$ of the denominator from Eq.(56) is found to be from the algebraic equation

$$(\Delta^4 - \omega^4)(\Delta^2 + \omega^2) = \frac{9\pi^2}{4}\omega^2\Delta^4 , \tag{57}$$

the approximate solution of which can be written in the form

$$\omega_{ext} \approx \frac{2\Delta}{\sqrt{9\pi^2 - 4}} \approx 0.217\Delta, \tag{58}$$

Thus, measuring the position of the maximum on the spectral density it is possible to determine the value of the superconducting energy gap. In fig. 10 the frequency dependences of the uniform spectral density of states $R_\sigma^N(\omega)$ from Eq.(56) in a normal metal with an induced gap for the superconducting energy gap value $\Delta=2.12 \cdot 10^{-4}$ eV and $\mu = 8.16$ eV with different barrier transparency coefficients $\Gamma_N$ are presented. From this figure it can be seen that with increasing $\Gamma_N$ the spectral density $R_\sigma^N(\omega)$ approaches zero that reflects the strengthening proximity effect, when a normal metal acquires the properties of a superconducting (see curve 5 in Fig. 10). On the other hand, with decrease barrier transparency the spectral density of states approaches closer to a purely metallic state (see straight line 1 in fig. 10). The behavior $R_\sigma^N(\omega)$ is consistent with experiment [14]. The differences obtained are due to only the fact that in this model the spatial dependence of both gap functions $\tilde{\Delta}$ and $\Delta$ does not take into account.

## 6. Conclusions

Summing up, it can be stated that the presented diagram method of time-perturbation theory in a zero approximation of the self-consistent field gives results that coincide with the approximation of the molecular field for a metal in the normal and superconducting states. Also, for the first time in the framework of this method the appearance of a proximity effect in the form of a complex energy gap induced by a superconductor in a normal metal is predicted. In the limit of the low barrier transparency an absolute value of the induced energy gap was found to be only a few percent of the superconducting one. The temperature and parametric dependences of the energy gap $|\tilde{\Delta}|$ as well as phase $\psi$ of induced order parameter are obtained. It is consistent with an increasing the barrier transparency and decreasing an energy gap of the superconductor with approaching the critical temperature. A sufficiently large value of the tunnel matrix element leads to the destruction of the Cooper pairs and absence of the proximity effect in the considered hybrid structure.

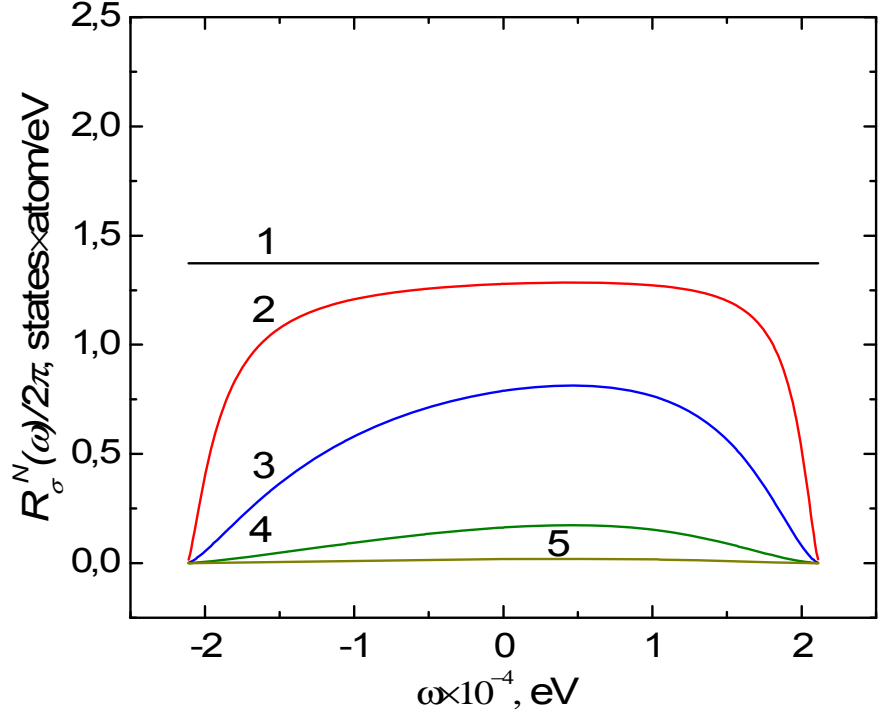

Fig. 10. The frequency dependences of the homogeneous spectral density of states $\frac{R_\sigma^N(\omega)}{2\pi}$ from Eq.(56) in a normal metal with induced gap $\tilde{\Delta}$ for superconducting energy gap value $\Delta=2.12\cdot 10^{-4}$ eV, chemical potential $\mu = 8.16$ eV and barrier transparency coefficients $\Gamma_N=0.$, $10^{-5}$, $10^{-4}$, $10^{-3}$ и $10^{-2}$ states· atom·eV(curves 1-5, respectively). The density of states at the Fermi level $\rho(\mu)=1.37$, states ·atom / eV is taken for Ti atom (4 electrons per atom).

With increasing a barrier transparency the spectral density decreases that reflects the proximity effect strengthening when a normal metal acquires superconducting properties. However, with a high transparency ($\delta r > \Delta$) the Cooper pairs in a normal metal are destroyed. When a barrier transparency decreases the density of states approaches a purely metallic one. The behavior is consistent with experiment although the existing quantitative differences are connected with the presence of the spatial dependence of the gap functions $\tilde{\Delta}$ and $\Delta$ that is not taken into account in the considered approximation. Also, this theory essentially simplifies a subsequent account for corrections due to the influence of fluctuations.

**Acknowledgements**

It is a pleasure to acknowledge a number of stimulating discussions with E.M. Rudenko and M.A.Belogolovsii. The study was carried out within the Fundamental Research Programme funded by the MES of Ukraine (Project No. 0117U002360).

**7. Appendix**

Let us calculate the integrals $\varphi(\omega)$ and $\chi(\omega)$ from Eq.(36). Consider a simpler case when $\mu \Box |\omega| > \Delta$. Then we have for the density of states $\rho(\varepsilon) = C\sqrt{\varepsilon}$ and $d(\omega) = \sqrt{\omega^2 - \Delta^2}$, where $C = \dfrac{V}{2\pi^2\hbar^3}(2m)^{3/2}$ is the constant for an electron with mass $m$ in a crystal with volume $V$.

$$\varphi(\omega) = -\int_0^\mu d\varepsilon \frac{C\sqrt{\varepsilon}}{(\varepsilon-\mu)^2 - d(\omega)^2} = -\frac{C}{2d(\omega)}\{L(\mu, d(\omega)) - L(\mu, -d(\omega))\} \approx \frac{\rho(\mu)}{4\mu^{3/2}}\ln\left(\frac{d(\omega)}{4\mu}\right) \sim 0$$

(A.1)

$$\chi(\omega) = -\int_0^\mu d\varepsilon \frac{C\sqrt{\varepsilon}(\varepsilon-\mu)}{(\varepsilon-\mu)^2 - d(\omega)^2} = -2C\sqrt{\mu} - \frac{C}{2}\{L(\mu, d(\omega)) + L(\mu, -d(\omega))\} \approx -\rho(\mu)\left\{2 + \ln\left(\frac{d(\omega)}{4\mu}\right)\right\},$$

(A.2)

where $L(\mu, d) = \sqrt{\mu+d}\ln\left|\dfrac{\sqrt{\mu+d}-\sqrt{\mu}}{\sqrt{\mu+d}+\sqrt{\mu}}\right|$. It is somewhat more difficult to calculate the indicated integrals with $|\omega| < \Delta$ when the function $b(\omega) = \sqrt{\Delta^2 - \omega^2}$ is real. But this case is the most important since the appearance of the proximity effect is connected with it. As a result, one can write

$$\varphi(\omega) = -\int_0^\mu d\varepsilon \frac{C\sqrt{\varepsilon}}{(\varepsilon-\mu)^2 + b(\omega)^2} = -\int_0^{\sqrt{\mu}} dt\, \frac{C}{p(\omega)}\left\{\frac{t}{t^2 - pt + q} - \frac{t}{t^2 + pt + q}\right\},$$

(A.3)

where

$$p^2(\omega) = 2(\mu + q)$$
$$q(\omega) = \sqrt{\mu^2 + b^2(\omega)}$$

(П.4)

Then it is easy to find that

$$\varphi(\omega) = -\frac{C}{2}\left\{\frac{1}{p(\omega)}\ln\left|\frac{\mu - p(\omega)\sqrt{\mu} + q(\omega)}{\mu + p(\omega)\sqrt{\mu} + q(\omega)}\right| + \frac{1}{a(\omega)}\arctan\left(\frac{\sqrt{\mu}}{a(\omega)}\right)\right\}, \qquad (A.5)$$

where $2a^2(\omega) = q(\omega) - \mu$. Since $b(\omega) \ll \mu$ we have an approximate equality

$$\varphi(\omega) \approx -\frac{1}{2}\rho(\mu)\left\{\frac{1}{\mu}\ln\frac{\sqrt{\Delta^2 - \omega^2}}{4\mu} + \frac{\pi}{2\sqrt{\Delta^2 - \omega^2}} - \frac{1}{\mu}\right\} \approx -\frac{\pi\rho(\mu)}{2\sqrt{\Delta^2 - \omega^2}} \qquad (A.6)$$

Simiraly, one can show that

$$\chi(\omega) \approx -\rho(\mu)\left\{2 + \ln\left(\frac{b(\omega)}{4\mu}\right)\right\} \qquad (A.7)$$

**References**

[1] A.A. Abrikosov, L.P. Gor'kov and I.Ye. Dzyaloshinskii. Quantum Field Theoretical Methods in Statistical Physics/ Pergamon Press: Oxford, 1965, 365 p.

[2] V. Chandrasekhar, Introduction to the quasiclassical theory of superconductivity for diffusive proximity-coupled systems. In: ThePhysics of Superconductors. V.2 (Springer-Verlag, 2004).

[3] Zh. Devizorova and S. Mironov, Phys.Rev. B **95**, 144514 (2017).

[4] J.A. Ouassou, A. Pal, M. Blamire, M. Eschrig, J. Linder, Scientific Reports. **16**, 7(1):1932 (2017).

[5] P.G. de Gennes, Rev. Mod. Phys. **36**, 225 (1964).

[6] W.L. McMillan. Phys.Rev. **175**, 537 (1968).

[7] L.P. Gor'kov. Sov.Phys.JETP. **34**(7), №.3, 505 (1958).

[8] G.M. Eliashberg. Sov.Phys.JETP. **12**, №.5, 1000 (1961).

[9] G. Eilenberger. Zeitschrift für Physik. **214**, 195 (1968).

[10] G.D. Mahan, Many Particle Physics (Plenum, New York, 1990), 793 p.

[11] J. Bardeen and J. Schrieffer. Recent Development in Superconductivity (Progress in Low Temperature Physics, V.III, Chapter VI). North-Holland Publishing Company, Amsterdam, 1961, 170 p.

[12] J.M. Luttinger and J.C. Ward. Phys.Rev. **118**, №.5, 1417 (1960).

[13] R. Meservey and B.B. Schwartz. Equilibrium Properties: Comparison of Experimental Results with Predictions of the BCS Theory (in book "Superconductivity" V.1, edited by R.D. Parks. Marcell Dekker: New York, 1969. 664 p.).

[14] S. Guéron, H. Pothier, Norman O. Birge, D. Esteve, and M.H. Devoret. Phys.Rev. Lett. **77**, 3025 (1996).